\documentclass[twocolumn]{aastex63}

\newcommand\cannon{{\em The Cannon }}

\received{August 4, 2021}
\accepted{January 31, 2020}
\submitjournal{AJ}

\shorttitle{OCCAM: V. New Cluster Abundances using {\it The Cannon}}
\shortauthors{Ray et al.}

\begin{document}

\title{The Open Cluster Chemical Abundances and Mapping Survey: V. \\ Chemical Abundances of CTIO/Hydra Clusters using \textit{The Cannon}}

\author{Amy E. Ray}
\affiliation{Department of Physics and Astronomy, Texas Christian University, TCU Box 298840 \\
Fort Worth, TX 76129, USA (a.e.ray, p.frinchaboy, j.donor, m.melendez@tcu.edu)}

\author[0000-0002-0740-8346]{Peter M. Frinchaboy}
\affiliation{Department of Physics and Astronomy, Texas Christian University, TCU Box 298840 \\
Fort Worth, TX 76129, USA (a.e.ray, p.frinchaboy, j.donor, m.melendez@tcu.edu)}

\author{John Donor}
\affiliation{Department of Physics and Astronomy, Texas Christian University, TCU Box 298840 \\
Fort Worth, TX 76129, USA (a.e.ray, p.frinchaboy, j.donor, m.melendez@tcu.edu)}

\author{S.~D.~Chojnowski} \affiliation{Department of Astronomy, New Mexico State University, Las Cruces, NM, 88001, USA}

\author{Matthew Melendez}
\affiliation{Department of Physics and Astronomy, Texas Christian University, TCU Box 298840 \\
Fort Worth, TX 76129, USA (a.e.ray, p.frinchaboy, j.donor, m.melendez@tcu.edu)}

\begin{abstract}

Open clusters are key chemical and age tracers of Milky Way evolution. While open clusters provide significant constraints on galaxy evolution, their use has been limited due to discrepancies in measuring abundances from different studies.  We analyze medium resolution ($R \sim 19,000$) CTIO/Hydra spectra of giant stars in 58 open clusters using \textit{The Cannon} to determine [Fe/H], [Mg/Fe], [Si/Fe], [Al/Fe], and [O/Fe]. This work adds an additional 55 primarily southern hemisphere open clusters calibrated to the SDSS/APOGEE DR16 metallicity system.  This uniform analysis is compared to previous studies [Fe/H] measurements for 23 clusters and we present spectroscopic metallicities for the first time for 35 open clusters.

\end{abstract}


\section{Introduction} \label{sec:intro}

Open clusters provide reliable ages, a key constraint needed to study Galactic chemical evolution since their stars formed at the same time out of similar material. Many studies over the past few decades \citep[e.g., ][]{janes_79,yong_2005,bragaglia2008,jacobson_2009,friel2010,carrera_2011,magrini_2017,casamiquela_2019,Donor18,Donor20} have used open clusters to explore the radial metallicity gradient in the Galactic disk, which has shown that clusters closer to the center are generally more metal rich than clusters in the outer Galaxy.

However, there are several problems in using open clusters to study the Galactic abundance gradient. One is the number of open clusters used in individual studies. A way to improve the current knowledge in this area is to increase the number of clusters with known chemical abundances. There are roughly 2000 known open clusters \citep{CG_20}, but only a small portion of them have been analyzed chemically. Even the ones with measured abundance values can have substantial uncertainties from study to study \citep{yong_2005}. A few reasons for such large uncertainties are due to varying data quality, the type of data, and different data analysis methods between studies. Another source of uncertainty arises depending on which catalog each survey chose for open cluster distances, as there are several that have determined substantially different distance results as discussed in \citet{Donor18}. This difference translates into widely varying results when attempting to determine a chemical abundance gradient across the disk of the Milky Way. \citet{Yong} and \citet{Donor18,Donor20} highlight this problem in their abundance gradient research. \citet{Metallicites_of_OC_1} compiled a homogenized sample of open cluster abundances, but there are still large uncertainties due to different types of observations and resolutions.

In this study, we put together a large uniform sample open clusters observed with CTIO/hydra spectra. For this work we are using \textit{The Cannon}, developed by \citet{Cannon}, which offers a unique way to find stellar parameters without having to use any models. Instead, this machine learning method takes a subset of stars with known parameters or ``labels" and creates a model based on pixel-to-pixel variations. This model can be applied to the rest of the set of stars to infer labels for them.
Our training set is based on the Apache Point Observatory Galactic Evolution Experiment Data Release 16 (APOGEE DR16) system in order to correct the problems with current surveys that were listed above. This sample trained with stars from \citet[][also on the APOGEE DR16 system]{Donor20} is designed to form a more extensive dataset for Galactic abundance studies. The sections in this paper are as follows: Section \ref{data} describes APOGEE DR16 and the observations taken for the sample of 58 open clusters. \textit{The Cannon} and the training set that we used are described in Section \ref{Cannon}. In Section \ref{results}, the results of this study are presented and we discuss comparison to similar studies, and finally conclusions in Section \ref{conclusion}.

\section{Data and Observations} \label{data}

The primary data for this study comes from optical data near the Calcium infrared triplet (7745--8730 \AA) taken with the Hydra spectrograph using the Cerro Tololo Inter-American Observatory (CTIO) 4m telescope.  Data was observed on UT 2002 March, 2003 March, 2003 July, and 2003 August.  The data reduction and radial velocity and membership analysis of this data was conducted using IRAF\footnote{IRAF is distributed by the National Optical Astronomy Observatory, which is operated by the Association of Universities for Research in Astronomy, Inc., under cooperative agreement with the National Science Foundation.} standard {\tt ccdproc}, {\tt dohydra}, and {\tt fxcor} routines, which is fully described in \citet{Frinchaboy}.  

We have chosen to train \cannon using data from the Sloan Digital Sky Survey’s \citep[SDSS;][]{Blanton,Eisenstein} sixteenth data release \citep[DR16;][]{DR16paperpaper,Holtzman18,Jonsson20} taken as part of the APOGEE \citep[APOGEE 2;][]{Majewski}. The APOGEE/DR16 dataset includes about 430,000 stars, collected using two APOGEE spectrographs \citep{wilson_2012} at Apache Point Observatory \citep[APO; New Mexico ][]{sloan_telescope} and Las Campanas Observatory \citep[LCO; Chile ][]{lco_telescope} .  For this study data from LCO provides a key overlap with our available training data.   The APOGEE data reduction pipeline \citep{DataReduc,Holtzman18} provides stellar atmospheric parameters and radial velocity measurements, while elemental abundances are provided from the ASPCAP pipeline \citep{APOGEEOutline,Holtzman18,Jonsson20}

\begin{figure*}[ht!]
\epsscale{1.35}
\plotone{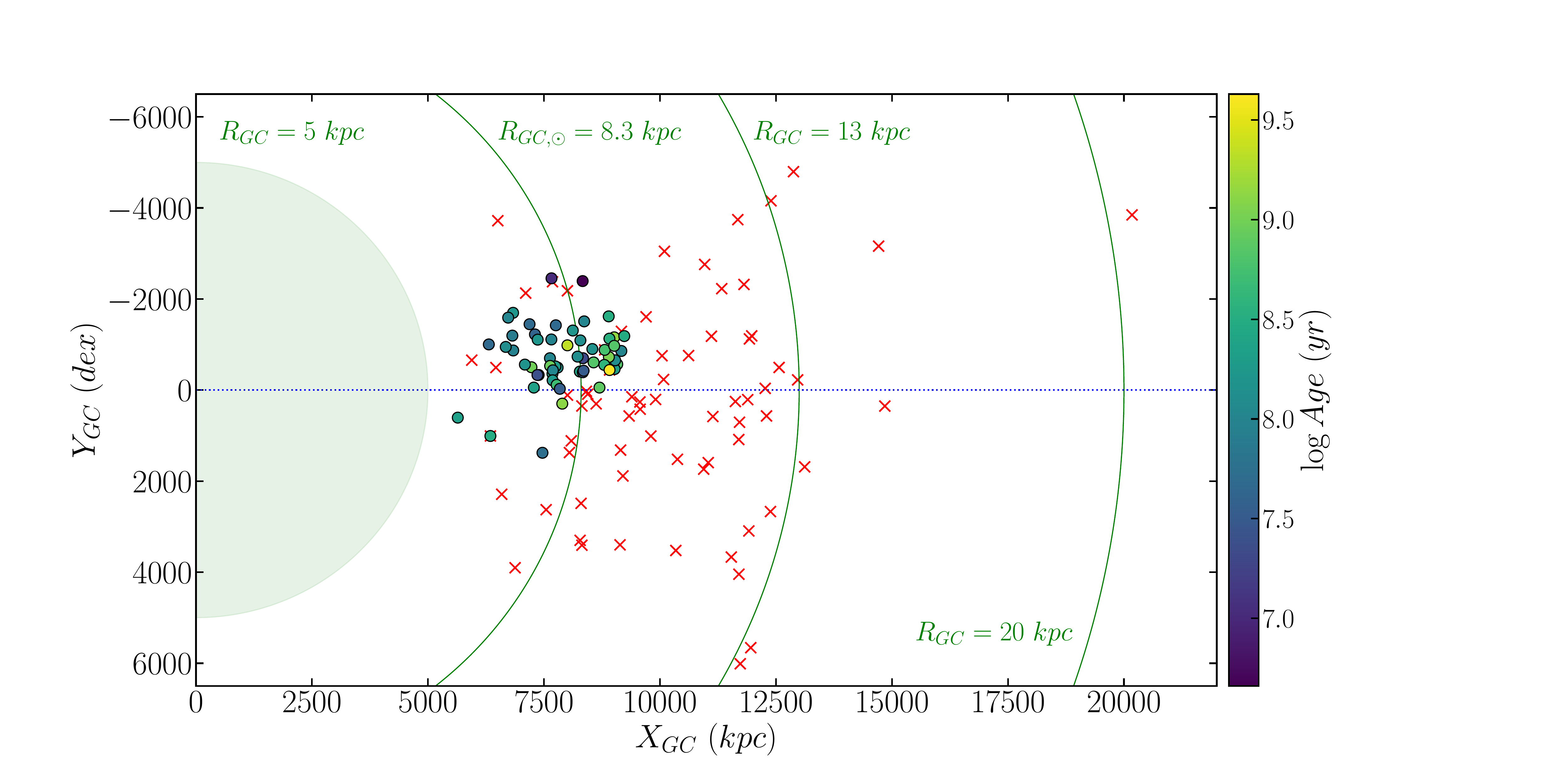}
    \caption{Spatial Galactic distribution of open clusters in \textit{this survey} color-coded with ages and $R_{GC}$ distance from \citet{CG_20}.  Also shown are the location of OCCAM-IV DR16-based cluster from \citet{Donor20} also updated with ages and $R_{GC}$ distance from \cite[][; red crosses]{CG_20}. 
    \label{fig:XY}}
\end{figure*}

For this study, we have applied \cannon to derive chemical abundances for selected available elements, after applying a cut based on the radial velocity cross-correlation quality Tonry-Davis Ratio \citep[$TDR \ge 11$][]{tdr}. After this cut we were able to use 25 stars from 3 clusters that are also observed with APOGEE DR16 in the training set.  The Galactic distribution of the clusters in our sample are shown in Figure \ref{fig:XY} using distances from \citep{CG_20} and also showing the positions of the high quality clusters from \citep{Donor20}.

Additionally, to verify the results from \citet{Frinchaboy}, we also used updated proper motion data from {\em Gaia} DR2 \citep{GaiaDR2} to re-check membership and found no change in membership between the Tycho-2 and Gaia-based proper motion selection.

In addition to the data collected as part of \citet{Frinchaboy}, observations of the the Sagittarius dwarf galaxy were also obtained on the same observing runs with the same instrument setup \citep{sgrpmf}.  Many of these stars were also observed as part of the APOGEE survey and so these additional 365 spectra are included in \cannon training set described below ($TDR \ge 11$). Data reduction and membership is discussed fully in \citet{sgrpmf}, which is the same as described in \citet{Frinchaboy}.
\cannon analysis of these Sagittarius data will be presented in Frinchaboy et al.\ ({\em in prep.}).

\section{\textit{The Cannon}} \label{Cannon}

\begin{figure*}[t!]
    \centering
    \epsscale{1.15}
    \plotone{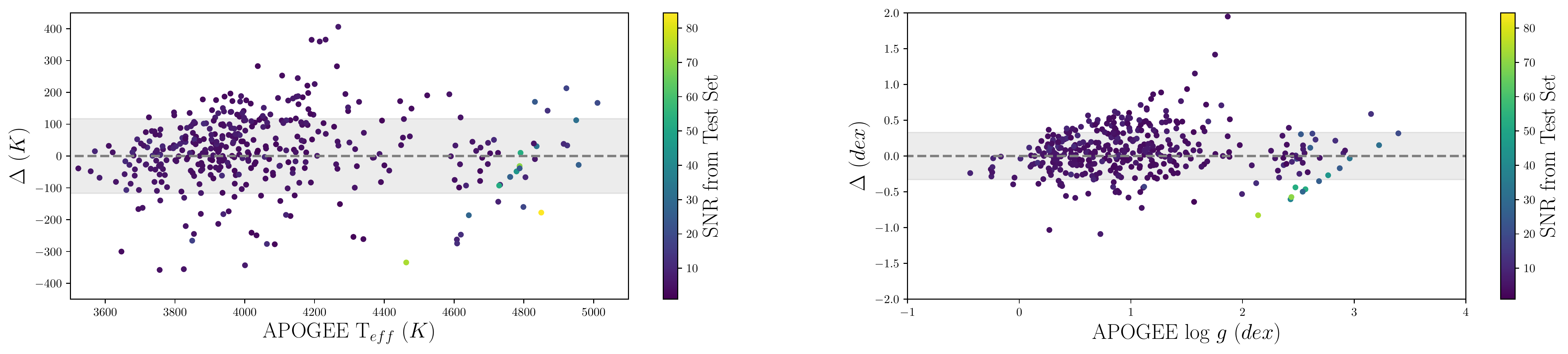}
    \caption{A comparison of effective temperature (T$_{eff}$) and surface gravity (log $g$) values obtained by \textit{The Cannon} and values obtained using APOGEE DR16 data. Most of the values fall within the scatter which is illustrated by the gray shaded areas.}
    \label{fig:train1}
\end{figure*}

\begin{figure}[h!]
    \centering
    \epsscale{1.05}
    \plotone{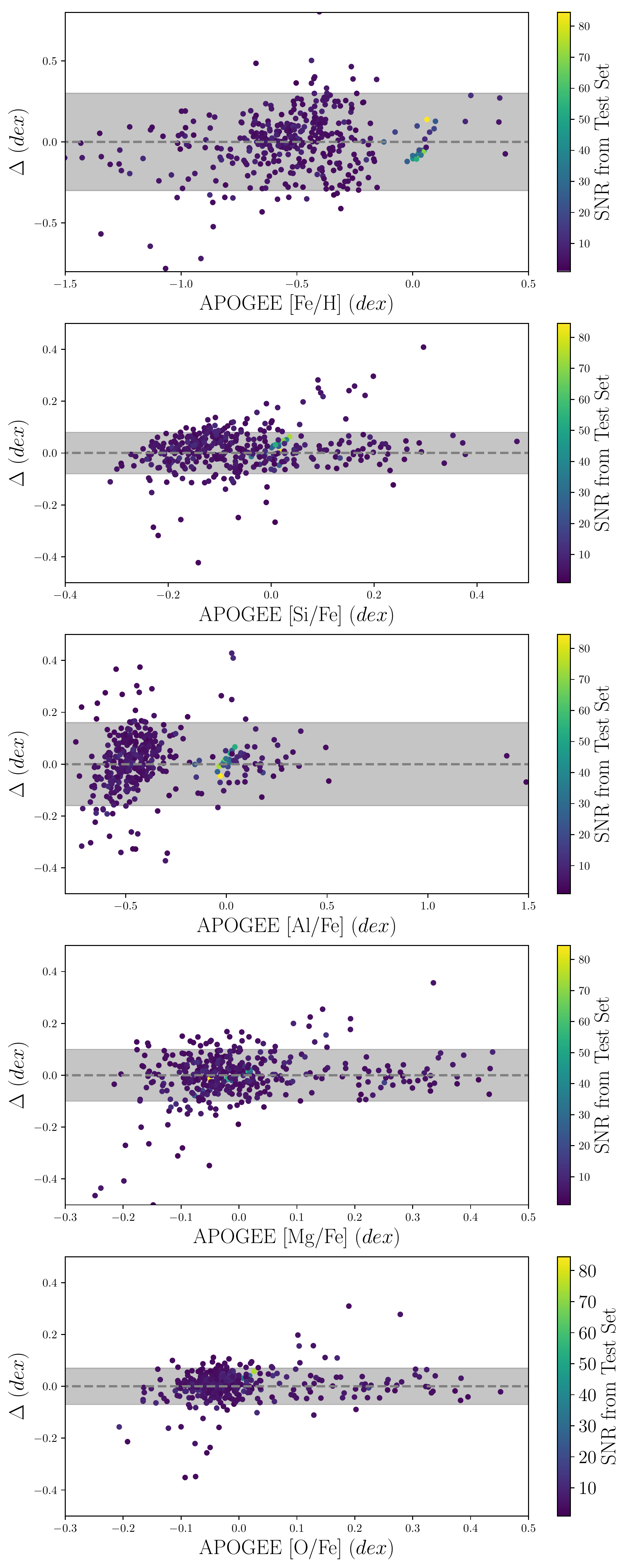}
    \caption{A comparison of [Fe/H], [Mg/Fe], [Si/Fe], [Al/Fe], and [O/Fe] values obtained by \textit{The Cannon} and values obtained using APOGEE DR16 data. The shaded areas represent the acceptable scatter as in \ref{fig:train1}. }
    \label{fig:train2}
\end{figure}

\subsection{Background}
Many high resolutions abundance studies use the curve of growth, the Boltzmann and Saha equations, and stellar models to determine chemical abundances. However, for this medium resolution, lower-S/N data from CTIO Hydra, we have chosen to use \textit{The Cannon} \citep{Cannon} which has been shown to work well for these types of data. 
The models, however, usually do not account for many distinctive factors in a star's atmosphere which means research groups can get opposing values for the same star \citep{Cannon}.  For this study, the set of parameters used were T$_{eff}$, $\log g$, [Fe/H], [Si/Fe], [Al/Fe], [Mg/Fe], and [O/Fe]  following the methodology from \citet{Ho17}.

\subsection{Training Set}

The training set was comprised of 390 stars in total with 25 stars from open clusters and 365 from the Sagittarius dwarf galaxy. When this set was used to train \textit{The Cannon}, a check was performed to determine if the output values were accurate. We first validated the output of the model by training The Cannon using 90\% of the training sample, then deriving labels for the remaining 10\%. We conducted this 10 times and analyze how the input labels compare to the output labels from this cross-validation test. 

Figure \ref{fig:train1} illustrates the full training set one-to-one plots for T$_{eff}$ and $\log g$, and Figure \ref{fig:train2} shows similar plots for [Fe/H], [Si/Fe], [Al/Fe], [Mg/Fe], and [O/Fe].  Values obtained from \textit{The Cannon} for the reference set are listed in Table \ref{tab:long} along with the values from APOGEE DR16. 
We find insignificant offset from DR16 for our parameters (Table \ref{tab:comp}), except maybe Si, but due to the lower signal-to-noise (S/N) for training set stars in the study, we obtain less accurate parameters in comparison to some larger studies \citep[e.g.,][]{Ho17, Sten20}.  Due to the small offset, and large scatter, we find that the errors are dominated by the scatter from low-S/N which are significantly larger that the uncertainties from the APOGEE DR1 survey input labels.

\begin{deluxetable*}{lrrrrrrrr}
\tablecaption{APOGEE DR16 Training Set Stars \label{tab:long} }
\tablecolumns{8}
\tablewidth{0pt}
\tablehead{
\colhead{2MASS ID} &
\colhead{S/N} &
\colhead{Temp} &
\colhead{$\log$ g} & 
\colhead{[Fe/H]} & 
\colhead{[Si/Fe]} & 
\colhead{[Al/Fe]} & 
\colhead{[Mg/Fe]} & 
\colhead{[O/Fe]} \\[-1.5ex] 
\colhead{} &
\colhead{(\small pixel$^{-1}$)} &
\colhead{(K)} &
\colhead{(dex)} & 
\colhead{(dex)} & 
\colhead{(dex)} & 
\colhead{(dex)} & 
\colhead{(dex)} & 
\colhead{(dex)}  
}
\startdata
2M18503032-2855221	&  9.2 & 3843 &	$-$0.4391	&	$-$0.8314	&	$-$0.2329	&	$-$0.5243	&	$-$0.1327	&	$-$0.1655	\\
2M19040855-3103228	&  4.6 & 3523 &	$-$0.2544	&	$-$0.5450	&	$-$0.2667	&	$-$0.4611	&	$-$0.0940	&	$-$0.1646	\\
2M18472966-2957462	&  7.7 & 3584 &	$-$0.2492	&	$-$0.5612	&	$-$0.1955	&	$-$0.3655	&	$-$0.0364	&	$-$0.0801	\\
2M18554328-2937095	&  5.7 & 3558 &	$-$0.2422	&	$-$0.6150	&	$-$0.1624	&	$-$0.4713	&	$+$0.0011	&	$-$0.1635	\\
2M18492810-2841275	&  9.0 & 3669 &	$-$0.2399	&	$-$0.6806	&	$-$0.1299	&	$-$0.5335	&	$-$0.0565	&	$-$0.0406	\\
2M18514730-2857148	&  9.5 & 3570 &	$-$0.2336	&	$-$0.5483	&	$-$0.1026	&	$-$0.3219	&	$-$0.0369	&	$-$0.0277	\\
2M18585375-3151470	&  7.3 & 3905 &	$-$0.1646	&	$-$0.7033	&	$-$0.1113	&	$-$0.4253	&	$+$0.0892	&	$-$0.0574	\\
2M18470776-3017243	&  5.6 & 3685 &	$-$0.0545	&	$-$0.5985	&	$-$0.1252	&	$-$0.5417	&	$+$0.0497	&	$-$0.0252	\\
2M18355640-2917489	&  5.8 & 3631 &	$-$0.0424	&	$-$0.7777	&	$-$0.1671	&	$-$0.4961	&	$-$0.0492	&	$-$0.0848	\\
2M18575903-3203421	& 10.2 & 3678 &	$+$0.0109	&	$-$0.2958	&	$-$0.1947	&	$-$0.4098	&	$-$0.1184	&	$-$0.1434	\\
  \multicolumn{8}{c}{\nodata} 
\enddata
\tablecomments{This table is available in its entirety in machine-readable form in the online journal. A portion is shown here for guidance regarding its form and content.}
\end{deluxetable*}

\begin{deluxetable}{lrr}
\tablecaption{APOGEE DR16/{\it The Cannon} Comparison for the Training Set\label{tab:comp} }
\tablecolumns{3}
\tablewidth{3in}
\tablehead{
\colhead{Parameter} &
\colhead{Mean Offset} &
\colhead{Std. Dev.} 
}
\startdata
$T_{eff}$     &  $+$17 K & 117 K \\ 
$\log$ g $\phantom{jkjkflljkljkl}$ &  $+$0.069 dex & $\phantom{jkljkjkl}$0.32 dex \\ 
{[Fe/H] }   & $+$0.001 dex & 0.27 dex \\ 
{[Si/Fe]}   & $+$0.010 dex & 0.09 dex \\ 
{[Al/Fe]}   & $+$0.004 dex & 0.16 dex \\ 
{[Mg/Fe]}   & $-$0.004 dex & 0.10 dex \\ 
{[O/Fe] }   & $+$0.004 dex & 0.07 dex \\ 
\enddata
\end{deluxetable}

\section{Results \& Discussion} \label{results}

Our resultant sample in this study consists of 58 open clusters with 237 member stars of which 35 have no previous spectroscopically determined metallicity measurements. The individual stellar abundance measurements for Fe as well as the $\alpha$-elements magnesium, silicon, and oxygen, plus the odd-$Z$ element aluminum are presented in Table \ref{tab:results1}.  Signal-to-noise (S/N) for all cluster stars is computed at 8000 \AA.  
We find that all of the elements scale relative to iron, in solar abundance ratios, as expected for clusters nearby the Sun and are within the normal of solar neighborhood cluster mean abundances as seen in \citet{Donor20}.
The resultant membership and clusters averages are for all clusters studied are presented in Table \ref{tab:results2}, which are calculated as in \citep{Donor20}.

\begin{deluxetable*}{lrrrrrrrrr}
\tablecaption{\cannon results star results for all clusters \label{tab:results1} }
\tablecolumns{9}
\tablewidth{0pt}
\tablehead{
\colhead{Cluster} &
\colhead{2MASS ID} &
\colhead{S/N} &
\colhead{Temp} &
\colhead{$\log$ g} & 
\colhead{[Fe/H]} & 
\colhead{[Si/Fe]} & 
\colhead{[Al/Fe]} & 
\colhead{[Mg/Fe]} & 
\colhead{[O/Fe]} \\[-1.5ex] 
\colhead{name} &
\colhead{} &
\colhead{(\small pixel$^{-1}$)} &
\colhead{(K)} &
\colhead{(dex)} & 
\colhead{(dex)} & 
\colhead{(dex)} & 
\colhead{(dex)} & 
\colhead{(dex)} & 
\colhead{(dex)}  
}
\startdata
           Collinder 205 &     09001428-4901040 & 47.4 &  5174 & 3.45 & $-$0.22 & $+$0.22 & $+$0.27 & $+$0.20 & $+$0.12  \\ 
           Collinder 205 &     09003203-4858199 & 28.0 &  4975 & 2.75 & $-$0.45 & $+$0.13 & $+$0.21 & $+$0.03 & $+$0.16  \\ \hline
           Collinder 258 &     12271397-6043435 & 14.5 &  4896 & 2.59 & $-$0.37 & $+$0.09 & $+$0.16 & $-$0.03 & $+$0.11  \\ 
           Collinder 258 &     12265230-6047561 & 20.1 &  5157 & 3.53 & $-$0.17 & $+$0.22 & $+$0.28 & $+$0.20 & $+$0.12  \\ 
           Collinder 258 &     12291805-6031347 & 70.1 &  5213 & 3.88 & $-$0.11 & $+$0.29 & $+$0.38 & $+$0.25 & $+$0.15  \\ \hline
              Harvard 10 &     16183313-5454573 & 47.6 &  4890 & 2.53 & $-$0.30 & $+$0.05 & $+$0.12 & $+$0.02 & $+$0.09  \\ 
              Harvard 10 &     16183125-5457397 & 40.3 &  5138 & 3.28 & $-$0.18 & $+$0.21 & $+$0.26 & $+$0.19 & $+$0.11  \\ 
              Harvard 10 &     16200896-5452364 & 21.3 &  4949 & 2.57 & $-$0.44 & $+$0.14 & $+$0.25 & $+$0.03 & $+$0.14  \\ 
              Harvard 10 &     16195235-5503168 & 21.8 &  4933 & 2.68 & $-$0.29 & $+$0.05 & $+$0.09 & $+$0.02 & $+$0.08  \\ 
              Harvard 10 &     16183216-5502259 & 44.9 &  5138 & 3.31 & $-$0.18 & $+$0.21 & $+$0.25 & $+$0.19 & $+$0.11  \\ 
  \multicolumn{9}{c}{\nodata} 
\enddata
\tablecomments{This table is available in its entirety in machine-readable form in the online journal. A portion is shown here for guidance regarding its form and content.}
\end{deluxetable*}

\begin{deluxetable*}{lrrrrrrrrrc}
\tablecaption{\cannon-based cluster parameters.  \label{tab:results2} }
\tabletypesize{\scriptsize}
\tablecolumns{9}
\tablewidth{0pt}
\tablehead{
\colhead{Cluster} &
\colhead{Num} & \colhead{$\log (Age)$\tablenotemark{a}} & \colhead{[Fe/H]} & 
\colhead{[Si/Fe]} & 
\colhead{[Al/Fe]} & 
\colhead{[Mg/Fe]} & 
\colhead{[O/Fe]} & 
 \colhead{New?}\\[-2ex]  
\colhead{name} & 
\colhead{Stars} & \colhead{(yrs)}&
\colhead{(dex)} & 
\colhead{(dex)} & 
\colhead{(dex)} & 
\colhead{(dex)} & 
\colhead{(dex)} & 
 & \colhead{}
}
\startdata
Collinder 205 	&   2 & 6.66 & $-$0.07 $\pm$   0.19 &  $+$0.02 $\pm$   0.06 &  $+$0.04 $\pm$   0.11 &  $-$0.03 $\pm$   0.07 &  $+$0.00 $\pm$   0.05   &  Y  \\ 
Collinder 258 	&   2 & 7.99 & $-$0.03 $\pm$   0.19 &  $+$0.02 $\pm$   0.06 &  $+$0.03 $\pm$   0.11 &  $-$0.03 $\pm$   0.07 &  $+$0.01 $\pm$   0.05   &  Y  \\ 
Harvard 10 	 	&   4 & 8.30 & $-$0.04 $\pm$   0.14 &  $+$0.02 $\pm$   0.05 &  $+$0.03 $\pm$   0.08 &  $-$0.03 $\pm$   0.05 &  $+$0.00 $\pm$   0.04   &  N  \\ 
IC 2395 	 	&   1 & 7.31 & $+$0.05 $\pm$   0.27 &  $+$0.01 $\pm$   0.09 &  $+$0.01 $\pm$   0.16 &  $+$0.01 $\pm$   0.10 &  $+$0.00 $\pm$   0.07   &  N  \\ 
IC 2488 	 	&   3 & 8.21 & $+$0.06 $\pm$   0.16 &  $+$0.02 $\pm$   0.05 &  $+$0.00 $\pm$   0.09 &  $-$0.01 $\pm$   0.06 &  $+$0.00 $\pm$   0.04   &  N  \\ 
IC 2581 	 	&   1 & 7.01 & $-$0.10 $\pm$   0.27 &  $-$0.01 $\pm$   0.09 &  $+$0.01 $\pm$   0.16 &  $-$0.05 $\pm$   0.10 &  $-$0.01 $\pm$   0.07   &  N  \\ 
IC 4651 	 	&   9 & 9.22 & $+$0.04 $\pm$   0.09 &  $+$0.00 $\pm$   0.03 &  $+$0.01 $\pm$   0.05 &  $+$0.01 $\pm$   0.03 &  $+$0.00 $\pm$   0.02   &  N  \\ 
IC 4756 	 	&   3 & 9.11 & $-$0.05 $\pm$   0.16 &  $+$0.06 $\pm$   0.05 &  $-$0.15 $\pm$   0.09 &  $+$0.02 $\pm$   0.06 &  $+$0.01 $\pm$   0.04   &  N  \\ 
Lynga 1 	 	&   2 & 8.22 & $-$0.01 $\pm$   0.19 &  $+$0.01 $\pm$   0.06 &  $+$0.02 $\pm$   0.11 &  $-$0.04 $\pm$   0.07 &  $+$0.00 $\pm$   0.05   &  N  \\ 
Lynga 2 	 	&   3 & 8.01 & $+$0.06 $\pm$   0.16 &  $+$0.03 $\pm$   0.05 &  $+$0.03 $\pm$   0.09 &  $+$0.01 $\pm$   0.06 &  $+$0.02 $\pm$   0.04   &  N  \\ 
NGC 1662 	 	&   3 & 8.89 & $+$0.04 $\pm$   0.16 &  $+$0.06 $\pm$   0.05 &  $-$0.07 $\pm$   0.09 &  $+$0.05 $\pm$   0.06 &  $+$0.04 $\pm$   0.04   &  N  \\ 
NGC 2215 	 	&   7 & 8.84 & $-$0.07 $\pm$   0.10 &  $+$0.02 $\pm$   0.03 &  $+$0.03 $\pm$   0.06 &  $-$0.02 $\pm$   0.04 &  $+$0.01 $\pm$   0.03   &  N  \\ 
NGC 2301 	 	&   3 & 8.33 & $+$0.09 $\pm$   0.16 &  $+$0.01 $\pm$   0.05 &  $+$0.02 $\pm$   0.09 &  $+$0.01 $\pm$   0.06 &  $+$0.01 $\pm$   0.04   &  N  \\ 
NGC 2323 	 	&   3 & 8.12 & $+$0.11 $\pm$   0.16 &  $+$0.01 $\pm$   0.05 &  $+$0.05 $\pm$   0.09 &  $+$0.03 $\pm$   0.06 &  $+$0.02 $\pm$   0.04   &  N  \\ 
NGC 2353 	 	&   5 & 8.01 & $-$0.16 $\pm$   0.12 &  $+$0.03 $\pm$   0.04 &  $+$0.04 $\pm$   0.07 &  $-$0.03 $\pm$   0.04 &  $+$0.02 $\pm$   0.03   &  N  \\ 
NGC 2354 	 	&  10 & 9.15 & $-$0.12 $\pm$   0.09 &  $+$0.02 $\pm$   0.03 &  $+$0.03 $\pm$   0.05 &  $-$0.03 $\pm$   0.03 &  $+$0.00 $\pm$   0.02   &  N  \\ 
NGC 2423 	 	&   8 & 9.04 & $+$0.10 $\pm$   0.10 &  $-$0.01 $\pm$   0.03 &  $-$0.01 $\pm$   0.06 &  $+$0.00 $\pm$   0.04 &  $-$0.02 $\pm$   0.02   &  N  \\ 
NGC 2437 	 	&   3 & 8.48 & $-$0.08 $\pm$   0.16 &  $+$0.02 $\pm$   0.05 &  $+$0.02 $\pm$   0.09 &  $-$0.02 $\pm$   0.06 &  $-$0.01 $\pm$   0.04   &  Y  \\ 
{\bf NGC 2447} 	&  17 & 8.76 & $-$0.06 $\pm$   0.07 &  $+$0.01 $\pm$   0.02 &  $+$0.02 $\pm$   0.04 &  $-$0.03 $\pm$   0.02 &  $-$0.02 $\pm$   0.02   &  {\bf T}\\ 
NGC 2482 	 	&   5 & 8.54 & $+$0.07 $\pm$   0.12 &  $-$0.01 $\pm$   0.04 &  $-$0.01 $\pm$   0.07 &  $+$0.01 $\pm$   0.04 &  $-$0.01 $\pm$   0.03   &  N  \\ 
NGC 2516 	 	&   3 & 8.38 & $-$0.09 $\pm$   0.16 &  $+$0.01 $\pm$   0.05 &  $+$0.05 $\pm$   0.09 &  $-$0.04 $\pm$   0.06 &  $-$0.01 $\pm$   0.04   &  N  \\ 
NGC 2527 	 	&   3 & 8.84 & $-$0.11 $\pm$   0.16 &  $+$0.02 $\pm$   0.05 &  $+$0.03 $\pm$   0.09 &  $-$0.04 $\pm$   0.06 &  $+$0.00 $\pm$   0.04   &  N  \\ 
NGC 2539 	 	&   9 & 8.84 & $-$0.08 $\pm$   0.09 &  $+$0.02 $\pm$   0.03 &  $+$0.02 $\pm$   0.05 &  $-$0.03 $\pm$   0.03 &  $-$0.01 $\pm$   0.02   &  N  \\ 
NGC 2546 	 	&   1 & 8.15 & $+$0.06 $\pm$   0.27 &  $+$0.01 $\pm$   0.09 &  $+$0.04 $\pm$   0.16 &  $+$0.04 $\pm$   0.10 &  $+$0.02 $\pm$   0.07   &  Y  \\ 
NGC 2547 	 	&   3 & 7.51 & $-$0.05 $\pm$   0.16 &  $+$0.02 $\pm$   0.05 &  $+$0.03 $\pm$   0.09 &  $-$0.02 $\pm$   0.06 &  $-$0.01 $\pm$   0.04   &  N  \\ 
NGC 2548 	 	&   3 & 8.59 & $+$0.12 $\pm$   0.16 &  $-$0.01 $\pm$   0.05 &  $+$0.00 $\pm$   0.09 &  $+$0.00 $\pm$   0.06 &  $-$0.02 $\pm$   0.04   &  N  \\ 
NGC 2567 	 	&   6 & 8.50 & $-$0.07 $\pm$   0.11 &  $+$0.01 $\pm$   0.04 &  $+$0.02 $\pm$   0.07 &  $-$0.03 $\pm$   0.04 &  $-$0.01 $\pm$   0.03   &  N  \\ 
NGC 2579 	 	&   2 & \nodata & $-$0.08 $\pm$   0.19 &  $+$0.01 $\pm$   0.06 &  $+$0.05 $\pm$   0.11 &  $-$0.03 $\pm$   0.07 &  $-$0.02 $\pm$   0.05   &  Y  \\ 
NGC 2669 	 	&   1 & 8.13 & $-$0.06 $\pm$   0.27 &  $+$0.00 $\pm$   0.09 &  $+$0.03 $\pm$   0.16 &  $-$0.04 $\pm$   0.10 &  $-$0.03 $\pm$   0.07   &  Y  \\ 
NGC 2670 	 	&   2 & 8.01 & $-$0.05 $\pm$   0.19 &  $+$0.03 $\pm$   0.06 &  $+$0.03 $\pm$   0.11 &  $-$0.02 $\pm$   0.07 &  $+$0.01 $\pm$   0.05   &  N  \\ 
{\bf NGC 2682} 	&  10 & 9.63 & $+$0.09 $\pm$   0.09 &  $-$0.02 $\pm$   0.03 &  $-$0.01 $\pm$   0.05 &  $+$0.00 $\pm$   0.03 &  $-$0.02 $\pm$   0.02   &  {\bf T}  \\ 
NGC 2925 	 	&   1 & 8.11 & $+$0.10 $\pm$   0.27 &  $+$0.02 $\pm$   0.09 &  $+$0.03 $\pm$   0.16 &  $+$0.01 $\pm$   0.10 &  $+$0.01 $\pm$   0.07   &  N  \\ 
NGC 3680 	 	&   6 & 9.34 & $-$0.05 $\pm$   0.11 &  $-$0.01 $\pm$   0.04 &  $+$0.01 $\pm$   0.07 &  $-$0.03 $\pm$   0.04 &  $-$0.02 $\pm$   0.03   &  N  \\ 
NGC 5138 	 	&   2 & 7.68 & $-$0.01 $\pm$   0.19 &  $+$0.02 $\pm$   0.06 &  $+$0.03 $\pm$   0.11 &  $-$0.01 $\pm$   0.07 &  $+$0.00 $\pm$   0.05   &  N  \\ 
NGC 5281 	 	&   7 & 7.60 & $-$0.02 $\pm$   0.10 &  $+$0.01 $\pm$   0.03 &  $+$0.02 $\pm$   0.06 &  $-$0.04 $\pm$   0.04 &  $-$0.01 $\pm$   0.03   &  Y  \\ 
NGC 5316 	 	&   4 & 8.22 & $-$0.01 $\pm$   0.14 &  $+$0.01 $\pm$   0.05 &  $+$0.03 $\pm$   0.08 &  $-$0.03 $\pm$   0.05 &  $-$0.02 $\pm$   0.04   &  N  \\ 
NGC 5460 	 	&   2 & 8.20 & $-$0.03 $\pm$   0.19 &  $+$0.02 $\pm$   0.06 &  $+$0.03 $\pm$   0.11 &  $-$0.02 $\pm$   0.07 &  $-$0.01 $\pm$   0.05   &  N  \\ 
NGC 5617 	 	&   6 & 8.02 & $-$0.04 $\pm$   0.11 &  $+$0.02 $\pm$   0.04 &  $+$0.03 $\pm$   0.07 &  $-$0.03 $\pm$   0.04 &  $+$0.00 $\pm$   0.03   &  N  \\ 
NGC 5662 	 	&   2 & 8.30 & $+$0.07 $\pm$   0.19 &  $+$0.03 $\pm$   0.06 &  $+$0.02 $\pm$   0.11 &  $+$0.00 $\pm$   0.07 &  $+$0.01 $\pm$   0.05   &  N  \\ 
NGC 5822 	 	&   3 & 8.96 & $+$0.01 $\pm$   0.16 &  $+$0.01 $\pm$   0.05 &  $+$0.02 $\pm$   0.09 &  $-$0.03 $\pm$   0.06 &  $-$0.02 $\pm$   0.04   &  N  \\ 
NGC 5823 	 	&   2 & 7.90 & $+$0.00 $\pm$   0.19 &  $+$0.02 $\pm$   0.06 &  $+$0.03 $\pm$   0.11 &  $-$0.03 $\pm$   0.07 &  $-$0.01 $\pm$   0.05   &  N  \\ 
NGC 6025 	 	&   3 & 8.02 & $+$0.19 $\pm$   0.16 &  $+$0.01 $\pm$   0.05 &  $+$0.04 $\pm$   0.09 &  $+$0.02 $\pm$   0.06 &  $+$0.00 $\pm$   0.04   &  N  \\ 
NGC 6031 	 	&   2 & 7.95 & $-$0.07 $\pm$   0.19 &  $+$0.02 $\pm$   0.06 &  $+$0.03 $\pm$   0.11 &  $-$0.03 $\pm$   0.07 &  $+$0.01 $\pm$   0.05   &  N  \\ 
NGC 6067 	 	&   1 & 8.10 & $+$0.03 $\pm$   0.27 &  $+$0.03 $\pm$   0.09 &  $+$0.06 $\pm$   0.16 &  $-$0.02 $\pm$   0.10 &  $+$0.02 $\pm$   0.07   &  N  \\ 
NGC 6124 	 	&  19 & 8.28 & $+$0.07 $\pm$   0.06 &  $+$0.02 $\pm$   0.02 &  $+$0.02 $\pm$   0.04 &  $+$0.00 $\pm$   0.02 &  $+$0.01 $\pm$   0.02   &  Y  \\ 
NGC 6134 		&   2 & 8.99 & $+$0.00 $\pm$   0.19 &  $+$0.01 $\pm$   0.06 &  $+$0.02 $\pm$   0.11 &  $-$0.02 $\pm$   0.07 &  $+$0.00 $\pm$   0.05   &  N  \\ 
NGC 6167 		&   3 & 8.19 & $+$0.05 $\pm$   0.16 &  $+$0.01 $\pm$   0.05 &  $+$0.02 $\pm$   0.09 &  $+$0.01 $\pm$   0.06 &  $+$0.01 $\pm$   0.04   &  Y  \\ 
NGC 6250 	 	&   1 & 7.38 & $-$0.15 $\pm$   0.27 &  $+$0.07 $\pm$   0.09 &  $-$0.17 $\pm$   0.16 &  $+$0.03 $\pm$   0.10 &  $+$0.00 $\pm$   0.07   &  Y  \\ 
NGC 6281 	 	&   7 & 8.71 & $-$0.06 $\pm$   0.10 &  $+$0.02 $\pm$   0.03 &  $+$0.03 $\pm$   0.06 &  $-$0.04 $\pm$   0.04 &  $+$0.00 $\pm$   0.03   &  N  \\ 
NGC 6405 	 	&   6 & 7.54 & $-$0.06 $\pm$   0.11 &  $+$0.02 $\pm$   0.04 &  $+$0.03 $\pm$   0.07 &  $-$0.03 $\pm$   0.04 &  $+$0.01 $\pm$   0.03   &  N  \\ 
NGC 6416 	 	&   6 & 8.36 & $-$0.07 $\pm$   0.11 &  $+$0.02 $\pm$   0.04 &  $+$0.03 $\pm$   0.07 &  $-$0.03 $\pm$   0.04 &  $+$0.01 $\pm$   0.03   &  N  \\ 
NGC 6603 	 	&   1 & 8.34 & $+$0.05 $\pm$   0.27 &  $-$0.01 $\pm$   0.09 &  $-$0.01 $\pm$   0.16 &  $-$0.04 $\pm$   0.10 &  $-$0.01 $\pm$   0.07   &  N  \\ 
{\bf NGC 6705} 	&   3 & 8.49 & $+$0.01 $\pm$   0.16 &  $+$0.06 $\pm$   0.05 &  $-$0.08 $\pm$   0.09 &  $+$0.01 $\pm$   0.06 &  $+$0.01 $\pm$   0.04   &  {\bf T}\\ 
NGC 6885 	 	&   2 & \nodata & $+$0.02 $\pm$   0.19 &  $+$0.06 $\pm$   0.06 &  $-$0.08 $\pm$   0.11 &  $+$0.01 $\pm$   0.07 &  $+$0.01 $\pm$   0.05   &  Y  \\ 
Roslund 3 	 	&   1 & 7.73 & $+$0.03 $\pm$   0.27 &  $+$0.01 $\pm$   0.09 &  $-$0.06 $\pm$   0.16 &  $+$0.01 $\pm$   0.10 &  $+$0.00 $\pm$   0.07   &  N  \\ 
Ruprecht 119 	&   2 & 7.68 & $+$0.08 $\pm$   0.19 &  $+$0.02 $\pm$   0.06 &  $+$0.03 $\pm$   0.11 &  $+$0.01 $\pm$   0.07 &  $+$0.01 $\pm$   0.05   &  Y  \\ 
Trumpler 10 	&   2 & 7.51 & $-$0.06 $\pm$   0.19 &  $+$0.02 $\pm$   0.06 &  $+$0.04 $\pm$   0.11 &  $-$0.02 $\pm$   0.07 &  $+$0.00 $\pm$   0.05   &  N  \\ 
Trumpler 18 	&   2 & 7.68 & $+$0.14 $\pm$   0.19 &  $+$0.01 $\pm$   0.06 &  $+$0.04 $\pm$   0.11 &  $+$0.03 $\pm$   0.07 &  $+$0.01 $\pm$   0.05   &  N  
\enddata
\tablenotetext{a}{Cluster ages taken from \citet{CG_20} }
\end{deluxetable*}

\subsection{Direct Comparison to \citet{Donor20}}

To verify our results, we first compare to previous work using APOGEE DR16 directly \citep{Donor20}.
We have 3 clusters in common with some stars in the \cannon training set.  While only a small subset of clusters, the studies agree within the uncertainties as shown in Table \ref{table:occam_comp} . 

\begin{deluxetable}{lccccc}[h!]
\tabletypesize{\scriptsize}
\tablecaption{Average in common open cluster iron abundance from \cite{Donor20} compared to \textit(this study).\label{table:occam_comp}}
	\tablehead{ \colhead{}& \multicolumn{2}{c}{\textit{This Study}} && \multicolumn{2}{c}{\textbf{Donor}}\\[.5ex]
\cline{2-3} \cline{5-6}
    \colhead{\textbf{Cluster}} &
    \colhead{\textbf{Number}} &
    \colhead{\textbf{[Fe/H]}} && 
    \colhead{\textbf{Number }} &
    \colhead{\textbf{[Fe/H]}}\\[-2ex]
    \colhead{\textbf{Name}} &
    \colhead{\textbf{of Stars}} &
    \colhead{\textbf{(dex)}} && 
    \colhead{\textbf{of Stars}} &
    \colhead{\textbf{(dex)}}}
\startdata
NGC 2447 &  17 & $-$0.06 $\pm$   0.07 && 3 & $-$0.08 $\pm$ 0.01   \\[.5ex]
NGC 2682 &  10 & $+$0.09 $\pm$   0.09 && 32 & $+$0.01 $\pm$ 0.03 \\[.5ex]
NGC 6705 &  3 & $+$0.01 $\pm$   0.16 && 12 & $+$0.12 $\pm$ 0.04 \\
\enddata
\end{deluxetable}


\subsection{Comparison to Other Surveys}

\begin{figure}[h!]
\epsscale{1.29}
\plotone{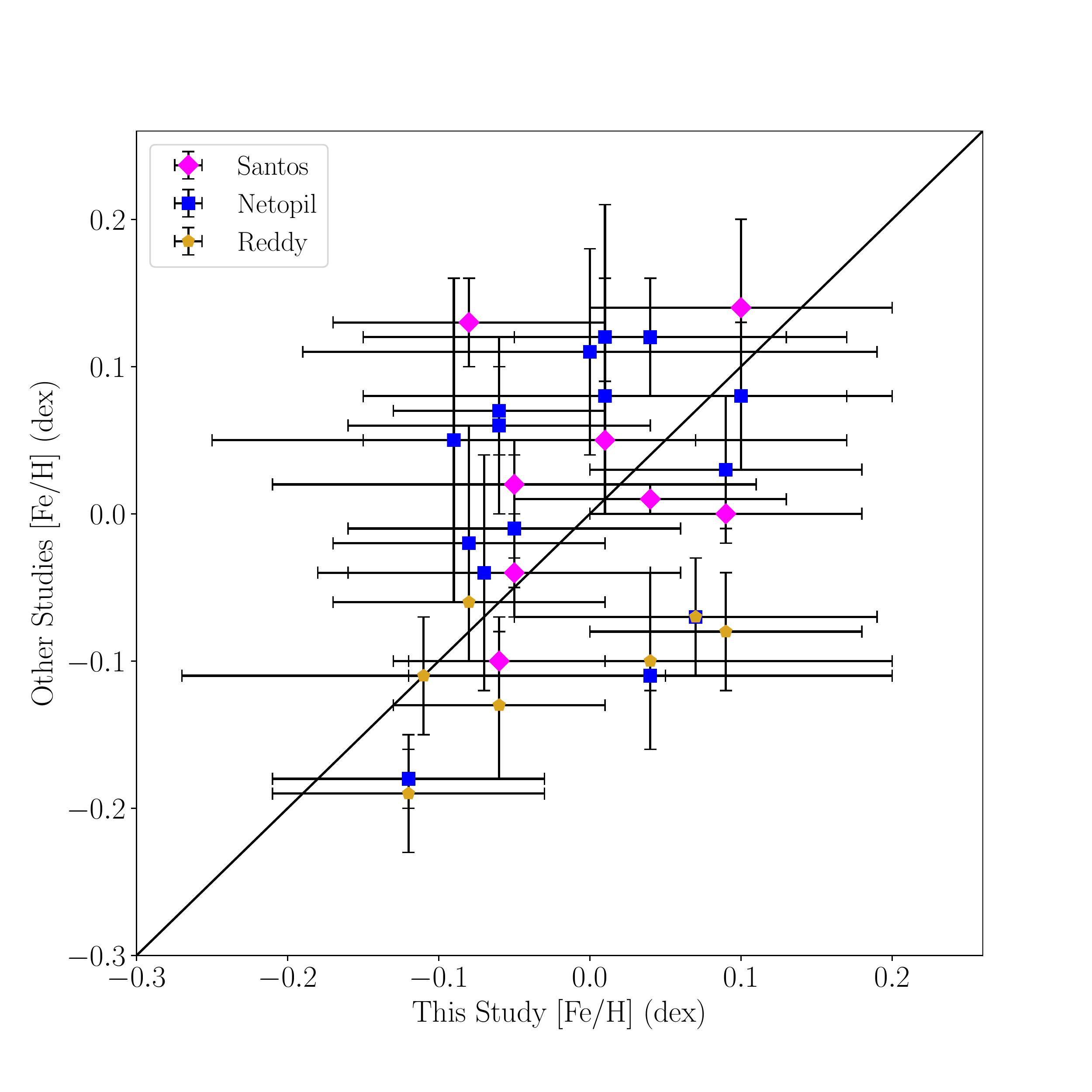}
    \caption{A comparison of average [Fe/H] values for in common open clusters from \textit{this survey} and from the literature compilation in Tables  \ref{table:Sample3}, \&  
    \ref{table:litrest}.  Magenta clusters are from \citet{Metallicites_of_OC_10},  blue clusters are from \citet{Metallicites_of_OC_1}, and orange clusters are from \citet{Metallicites_of_OC_14,Metallicites_of_OC_2}.
    \label{fig:NRSComp}}
\end{figure}

The \cite{Metallicites_of_OC_10} survey provided iron abundances for 13 open clusters using high-resolution spectra. There were 8 clusters that overlapped with this survey, and their values are listed in Table \ref{table:litrest}. To better illustrate how the two surveys compare, we constructed a one-to-one plot of iron abundances from \textit{The Cannon} vs. values from three significant studies, including \cite{Metallicites_of_OC_10}, which is shown in Figure \ref{fig:NRSComp}. All of the cluster values from \cite{Metallicites_of_OC_10} lie within the average uncertainties, although our values were slightly more metal-poor.

The next comparison was to the \citet{Metallicites_of_OC_14} and \citet{Metallicites_of_OC_2} studies. Both examined a total of 12 clusters using high-resolution spectra. Here, there was also an overlap of 7 clusters between this survey and the combined Reddy surveys. The values for each are shown in Table \ref{table:litrest}. These clusters are also shown on the one-to-one plot Figure \ref{fig:NRSComp}. The values obtained by \citet{Metallicites_of_OC_14,Metallicites_of_OC_2} were more metal-poor than the values determined using \textit{The Cannon}. 
\citet{Metallicites_of_OC_14} and \citet{Metallicites_of_OC_2} have been found to be slightly more metal-poor with respect to most measurements other high-resolution measurements, as summarized in \citet{Donor18} and seen in Figure 2 from \citet{Reddy16} in their comparison to previous literature results.

The last spectroscopic survey compared to was by \citet{Metallicites_of_OC_1}. They examined 172 clusters with a variety of data including low, medium, and high-resolution spectra as well as photometric data. The 12 clusters that overlapped had iron abundances determined from high-resolution spectra. All of the iron abundances are listed in Table \ref{table:litrest} and the one-to-one plot Figure \ref{fig:NRSComp}, which shows that the majority of the iron abundances are consistent and most clusters lie within the scatter range from \textit{The Cannon}. There are two outliers which are discussed in \S \ref{diff}.

\begin{deluxetable*}{lrrrrlr}[h!]
\tabletypesize{\scriptsize}
\tablecaption{Average open cluster iron abundance for open clusters in common between \textit{this study} and literature studies.\label{table:litrest}  }
	\tablehead{ \colhead{} & \multicolumn{2}{c}{\textit{This Study}} && \multicolumn{2}{c}{\textbf{Other Studies}}\\
    \cline{2-3} \cline{5-6}
    \colhead{\textbf{Cluster}} &
    \colhead{\textbf{Stars}} &
    \colhead{\textbf{ [Fe/H] }} &&
    \colhead{\textbf{Stars}} &
    \colhead{\textbf{ [Fe/H] }} &
    \colhead{\textbf{Citation}} \\[-2.5ex]
    \colhead{\textbf{Name}} &
    \colhead{\textbf{}} &
    \colhead{\textbf{(dex)}} && 
    \colhead{\textbf{}} &
    \colhead{\textbf{(dex)}} &
    \colhead{\textbf{}} }
\startdata
IC 2581    &  1 & $-$0.10 $\pm$ 0.27  &&  1  &  $-$0.34 $\pm$ \nodata & \cite{1994ApJS...91..309L} \\[.2ex]\hline
IC 4651    & 9  & $+$0.04 $\pm$ 0.09  &&  3  &  $+$0.01 $\pm$ 0.01 & \cite{Metallicites_of_OC_10}  \\[.2ex]
           &    &                     && 18  &  $+$0.12 $\pm$ 0.04 & \cite{Metallicites_of_OC_1}   \\[.2ex]
           &    &                     &&  5  &  $+$0.12 $\pm$ 0.05 & \citet{2008AA...489..403P}.   \\[.2ex]
           &    &                     &&  3  &  $+$0.11 $\pm$ 0.01 & \citet{Metallicites_of_OC_12} \\[.2ex]\hline
IC 4756    & 3  & $-$0.05 $\pm$ 0.16  &&  3  &  $+$0.02 $\pm$ 0.02 & \cite{Metallicites_of_OC_10}  \\[.2ex]
           &    &                     &&  9  &  $-$0.02 $\pm$ 0.01 & \citet{Metallicites_of_OC_3}  \\[.2ex]
           &    &                     && 12  &  $-$0.01 $\pm$ 0.10 & \citet{Metallicites_of_OC_4}  \\[.2ex]
           &    &                     &&  2  &  $+$0.01 $\pm$ \nodata& \cite{Metallicites_of_OC_5} \\[.2ex]\hline
NGC 1662   & 3  & $+$0.04 $\pm$ 0.16  &&  2  &  $-$0.10 $\pm$ 0.06 & \cite{Metallicites_of_OC_14,Metallicites_of_OC_2} \\[.2ex]
           &    &                     &&  2  &  $-$0.11 $\pm$ 0.01 & \cite{Metallicites_of_OC_1}.  \\[.2ex] \hline
NGC 2354   & 10 & $-$0.12 $\pm$ 0.09  &&  2  &  $-$0.19 $\pm$ 0.04 & \cite{Metallicites_of_OC_14,Metallicites_of_OC_2} \\[.2ex]
           &   &                      &&  2  &  $-$0.18 $\pm$ 0.02 & \cite{Metallicites_of_OC_1}   \\[.2ex] \hline
NGC 2423   & 8  & $+$0.10 $\pm$ 0.10  &&  3  &  $+$0.14 $\pm$ 0.06 & \cite{Metallicites_of_OC_10}  \\[.2ex]
           &    &                     &&  3  &  $+$0.08 $\pm$ 0.05 & \cite{Metallicites_of_OC_1}   \\[.2ex] \hline
NGC 2447   & 17 & $-$0.06 $\pm$ 0.07  &&  3  &  $-$0.08 $\pm$ 0.01 & \cite{Donor20}                \\[.2ex]
           &    &                     &&  3  &  $-$0.10 $\pm$ 0.03 & \cite{Metallicites_of_OC_10}  \\[.2ex]
           &    &                     &&  3  &  $-$0.13 $\pm$ 0.05 & \cite{Metallicites_of_OC_14,Metallicites_of_OC_2} \\[.2ex]
           &    &                     &&  4  &  $+$0.07 $\pm$ 0.03 & \cite{Metallicites_of_OC_1}   \\[.2ex] 
           &    &                     && 12  &  $-$0.17 $\pm$ 0.05 & \citet{2018MNRAS.476.4907D}   \\[.2ex] \hline
NGC 2482   & 5  & $+$0.07 $\pm$ 0.12  &&  1  &  $-$0.07 $\pm$ 0.04 & \cite{Metallicites_of_OC_14,Metallicites_of_OC_2} \\[.2ex]
           &    &                     &&  1  &  $-$0.07 $\pm$ \nodata & \cite{Metallicites_of_OC_1}\\[.2ex] \hline
NGC 2527   & 3  & $-$0.11 $\pm$ 0.16  &&  2  &  $-$0.11 $\pm$ 0.04 & \cite{Metallicites_of_OC_14,Metallicites_of_OC_2} \\[.2ex]\hline
NGC 2516   & 3  & $-$0.09 $\pm$ 0.16  &&  2  &  $+$0.05 $\pm$ 0.11 & \cite{Metallicites_of_OC_1}   \\[.2ex] \hline
NGC 2539   & 9  & $-$0.08 $\pm$ 0.09  &&  3  &  $+$0.13 $\pm$ 0.03 & \cite{Metallicites_of_OC_10}  \\[.2ex]
           &    &                     &&  2  &  $-$0.06 $\pm$ 0.04 & \cite{Metallicites_of_OC_14,Metallicites_of_OC_2} \\[.2ex]
           &    &                     &&  4  &  $-$0.02 $\pm$ 0.08 & \cite{Metallicites_of_OC_1}   \\[.2ex]
           &    &                     && 12  &  $-$0.03 $\pm$ 0.07 & \citet{2020MNRAS.494.1470M}   \\[.2ex]\hline
NGC 2548   &  3 & $+$0.12 $\pm$ 0.16  && 95  &  $-$0.06 $\pm$ 0.01 & \citet{2020AJ....159..220S}   \\[.2ex]\hline
NGC 2567   & 6  & $-$0.07 $\pm$ 0.11  &&  3  &  $-$0.04 $\pm$ 0.08 & \cite{Metallicites_of_OC_1}   \\[.2ex] \hline
NGC 2682   & 10 & $+$0.09 $\pm$ 0.09  && 32  &  $+$0.01 $\pm$ 0.03 & \cite{Donor20}                \\[.2ex]
           &    &                     &&  3  &  $+$0.00 $\pm$ 0.01 & \cite{Metallicites_of_OC_10}  \\[.2ex]
           &    &                     &&  3  &  $-$0.08 $\pm$ 0.04 & \cite{Metallicites_of_OC_14,Metallicites_of_OC_2} \\[.2ex]
           &    &                     && 27  &  $+$0.03 $\pm$ 0.05 & \cite{Metallicites_of_OC_1}   \\[.2ex]
           &    &                     &&  6  &  $+$0.03 $\pm$ 0.04 & \citet{2008AA...489..403P}    \\[.2ex]\hline
NGC 3680   & 6  & $-$0.05 $\pm$ 0.11  &&  3  &  $-$0.04 $\pm$ 0.01 & \cite{Metallicites_of_OC_10}  \\[.2ex]
           &    &                     && 10  &  $-$0.01 $\pm$ 0.06 & \cite{Metallicites_of_OC_1}   \\[.2ex]
           &    &                     &&  6  &  $-$0.06 $\pm$ 0.07 & \citet{2018ApJ...854..184P}   \\[.2ex]
           &    &                     && 11  &  $-$0.03 $\pm$ 0.02 & \citet{2012MNRAS.422.3527M}   \\[.2ex]
           &    &                     &&  2  &  $+$0.04 $\pm$ 0.03 & \citet{2008AA...489..403P}    \\[.2ex]\hline
NGC 5617   &  6 & $-$0.04 $\pm$ 0.11  &&  2  &  $-$0.18 $\pm$ 0.02 & \citet{Metallicites_of_OC_7}  \\[.2ex]\hline
NGC 5822   & 3  & $+$0.01 $\pm$ 0.16  &&  3  &  $+$0.05 $\pm$ 0.04 & \cite{Metallicites_of_OC_10}  \\[.2ex]
           &    &                     &&  7  &  $+$0.08 $\pm$ 0.08 & \cite{Metallicites_of_OC_1}   \\[.2ex]
           &    &                     && 11  &  $-$0.09 $\pm$ 0.06 & \citet{2018ApJ...854..184P}   \\[.2ex]
           &    &                     &&  3  &  $+$0.15 $\pm$ 0.08 & \citet{2010AA...515A..28P}    \\[.2ex]
           &    &                     &&  7  &  $+$0.08 $\pm$ 0.08 & \citet{1994ApJS...91..309L}   \\[.2ex]\hline
NGC 6067   &  1 & $+$0.03 $\pm$ 0.27  &&  5  &  $+$0.19 $\pm$ 0.05 & \citet{Metallicites_of_OC_11} \\[.2ex]\hline
NGC 6134   & 2  & $+$0.00 $\pm$ 0.19  &&  8  &  $+$0.11 $\pm$ 0.07 & \cite{Metallicites_of_OC_1}   \\[.2ex] 
           &    &                     &&  6  &  $+$0.15 $\pm$ 0.07 & \citet{2004AA...422..951C}    \\[.2ex]\hline
NGC 6281   & 7  & $-$0.06 $\pm$ 0.10  &&  2  &  $+$0.06 $\pm$ 0.06 & \cite{Metallicites_of_OC_1}   \\[.2ex]\hline
NGC 6405   &  6 & $-$0.06 $\pm$ 0.11  && 44  &  $+$0.07 $\pm$ 0.03 & \citet{Metallicites_of_OC_23} \\[.2ex]\hline
NGC 6603   &  1 & $+$0.05 $\pm$ 0.27  &&  7  &  $+$0.34 $\pm$ 0.15 & \citet{2015AA...578A..27C}    \\[.2ex]\hline
NGC 6705   &  3 & $+$0.01 $\pm$ 0.16  && 12  &  $+$0.12 $\pm$ 0.04 & \cite{Donor20}                \\[.2ex]
           &    &                     && 21  &  $+$0.12 $\pm$ 0.09 & \cite{Metallicites_of_OC_1}   \\[.2ex]
\enddata

\end{deluxetable*}

The remaining clusters with iron abundances from smaller high-resolution studies were also examined for inconsistencies. Table \ref{table:litrest} lists all of the values for [Fe/H] determined by \textit{The Cannon}, the [Fe/H] values from the literature, and the type of data that was used.  Most of the clusters fall within the range of scatter again (Table \ref{table:litrest}); however, 3 clusters were not. Reasons for why these values do not appear to agree are discussed in \S \ref{diff}.

The cluster NGC 2682 is one of the most well studied open clusters, therefore it was used as a calibration cluster. It also has an average [Fe/H] value determined from APOGEE DR16 data from \citet{Donor20} making it a significant check on how well \textit{The Cannon} produced values. The average [Fe/H] from \citet{Donor20} was 0.01$\pm$0.03, and this survey found a value of 0.09 $\pm$ 0.16. Additionally, the five other studies compared to in this paper with [Fe/H] determined for NGC 2682 \citep{Metallicites_of_OC_10,Metallicites_of_OC_14,Metallicites_of_OC_2,Metallicites_of_OC_1,2008AA...489..403P} agree within the uncertainties.

\subsection{Discrepancies}\label{diff}

There were 3 clusters that were almost outside of the acceptable scatter when compared to the literature values, which are significant given our larger uncertainties  
due to the lower S/N of this study. 
This section discusses possible reasons why there were differences in iron abundance values.

The value obtained for NGC 6705 was within the uncertainty of \citet{Donor20}, however it was more metal-poor. This is likely due to \citet{Donor20} having four times the number of member stars used for average abundance analysis as well as higher resolution, higher-S/N spectra than this study. The same is also true when compared to the results of \citep{Metallicites_of_OC_1}

For NGC 1662, both \citet{2015MNRAS.450.4301R} and \citet{Metallicites_of_OC_1} had values that were in agreement and for reference, these values are listed in Tables 6 and 7. The value for NGC 1662 from this study was more metal-rich than both \citet{2015MNRAS.450.4301R} and \citet{Metallicites_of_OC_1} and the comparison is shown in Figure 4. Each of these studies used high-resolution spectra of two stars where we used medium-resolution spectra of three stars.

The cluster NGC 2482 showed similar discrepancies to NGC 1662. \citet{2013MNRAS.431.3338R} and \citet{Metallicites_of_OC_1} found more metal-poor values for this cluster and each only used one star for the average abundance analysis compared to five stars used in this study, therefore cluster membership of the other studies may be the reason for such a large offset. 
This comparison is also shown in Figure \ref{fig:NRSComp}.

\section{Conclusion} \label{conclusion}

Using CTIO/Hydra medium-resolution ($R \sim 19,000$) spectroscopy of clusters stars with RV and proper motion membership determinations,  
we determined the mean [Fe/H] values for a set of 58 open clusters using \textit{The Cannon}. 

With this study:
\begin{itemize}
    \item We measured the first spectroscopic metallicity [Fe/H] for 35 open clusters with member stars verified by \citet{Frinchaboy} and re-verified with updated proper motions from \citet{GaiaDR2}. 
    \item We confirm the overall abundance scale, based on APOGEE DR16 \citep{DR16paperpaper} is generally consistent with other study for the 22 clusters that have spectroscopic metallicity determinations. 
    \item We find the clusters in this study have abundance ratios for oxygen, silicon, magnesium, and aluminum consistent with solar values, which is reasonable for clusters near the solar neighborhood.
    
\end{itemize}
These clusters add to the work of \citet{Donor20} yielding a combined dataset of over 150 clusters, all on the APOGEE DR16 abundance scale.  

\begin{acknowledgements}

We would like to thank Anna Ho for help in assisting in setting up \cannon for this work.
A.E.R, P.M.F., J.D., and M.M.\ acknowledge support for this research from the National Science Foundation (AST-1311835 \& AST-1715662).  P.M.F.\ also acknowledges some of this was performed at the Aspen Center for Physics, which is supported by National Science Foundation grant PHY-1607611.  P.M.F. also acknowledges travel support from NOAO/NOIRLab from the original collection of this data.

Funding for the Sloan Digital Sky Survey IV has been provided by the Alfred P. Sloan Foundation, the U.S. Department of Energy Office of Science, and the Participating Institutions. SDSS-IV acknowledges
support and resources from the Center for High-Performance Computing at
the University of Utah. The SDSS web site is www.sdss.org.

SDSS-IV is managed by the Astrophysical Research Consortium for the 
Participating Institutions of the SDSS Collaboration including the 
Brazilian Participation Group, the Carnegie Institution for Science, 
Carnegie Mellon University, the Chilean Participation Group, the French Participation Group, Harvard-Smithsonian Center for Astrophysics, 
Instituto de Astrof\'isica de Canarias, The Johns Hopkins University, 
Kavli Institute for the Physics and Mathematics of the Universe (IPMU) / 
University of Tokyo, Lawrence Berkeley National Laboratory, 
Leibniz Institut f\"ur Astrophysik Potsdam (AIP),  
Max-Planck-Institut f\"ur Astronomie (MPIA Heidelberg), 
Max-Planck-Institut f\"ur Astrophysik (MPA Garching), 
Max-Planck-Institut f\"ur Extraterrestrische Physik (MPE), 
National Astronomical Observatories of China, New Mexico State University, 
New York University, University of Notre Dame, 
Observat\'ario Nacional / MCTI, The Ohio State University, 
Pennsylvania State University, Shanghai Astronomical Observatory, 
United Kingdom Participation Group,
Universidad Nacional Aut\'onoma de M\'exico, University of Arizona, 
University of Colorado Boulder, University of Oxford, University of Portsmouth, 
University of Utah, University of Virginia, University of Washington, University of Wisconsin, 
Vanderbilt University, and Yale University.

This work has made use of data from the European Space Agency (ESA) mission {\it Gaia} (\url{https://www.cosmos.esa.int/gaia}), processed by the {\it Gaia} Data Processing and Analysis Consortium (DPAC, \url{https://www.cosmos.esa.int/web/gaia/dpac/consortium}). Funding for the DPAC has been provided by national institutions, in particular the institutions participating in the {\it Gaia} Multilateral Agreement.

This research made use of Astropy, a community-developed core Python package for Astronomy (Astropy Collaboration, 2018).
\end{acknowledgements}

\facilities{ Sloan (APOGEE), FLWO:2MASS, {\it Gaia}, CTIO:Hydra}. 
\software{\href{http://www.astropy.org/}{Astropy}, \href{https://github.com/annayqho/TheCannon}{\cannon}}

\bibliography{Ray.bib}

\end{document}